\documentclass[preprint,aps,prc,tightenlines,nofootinbib,showpacs]{revtex4}
\begin{document}
\input{psfig}
\input{epsf}
\def\Im{\mbox{\sl Im\ }}
\def\pd{\partial}
\def\oln{\overline}
\def\olft{\overleftarrow}
\def\ds{\displaystyle}
\def\bgreek#1{\mbox{\boldmath $#1$ \unboldmath}}
\def\sla#1{\slash \hspace{-2.5mm} #1}
\newcommand{\bra}{\langle}
\newcommand{\ket}{\rangle}
\newcommand{\vep}{\varepsilon}
\newcommand{\met}{{\mbox{\scriptsize met}}}
\newcommand{\lab}{{\mbox{\scriptsize lab}}}
\newcommand{\cm}{{\mbox{\scriptsize cm}}}
\newcommand{\mcal}{\mathcal}
\newcommand{\Del}{$\Delta$}
\newcommand{\g}{{\rm g}}
\long\def\Omit#1{}
\long\def\omit#1{\small #1}
\def\beq{\begin{equation}}
\def\eeq{\end{equation} }
\def\bea{\begin{eqnarray}}
\def\eea{\end{eqnarray}}
\def\eqref#1{Eq.~(\ref{eq:#1})}
\def\eqlab#1{\label{eq:#1}}
\def\figref#1{Fig.~\ref{fig:#1}}
\def\figlab#1{\label{fig:#1}}
\def\tabref#1{Table \ref{tab:#1}}
\def\tablab#1{\label{tab:#1}}
\def\secref#1{Section~\ref{sec:#1}}
\def\seclab#1{\label{sec:#1}}
\def\VYP#1#2#3{{\bf #1}, #3 (#2)}  % Volume, page (Year)
\def\NP#1#2#3{Nucl.~Phys.~\VYP{#1}{#2}{#3}}
\def\NPA#1#2#3{Nucl.~Phys.~A~\VYP{#1}{#2}{#3}}
\def\NPB#1#2#3{Nucl.~Phys.~B~\VYP{#1}{#2}{#3}}
\def\PL#1#2#3{Phys.~Lett.~\VYP{#1}{#2}{#3}}
\def\PLB#1#2#3{Phys.~Lett.~B~\VYP{#1}{#2}{#3}}
\def\PR#1#2#3{Phys.~Rev.~\VYP{#1}{#2}{#3}}
\def\PRC#1#2#3{Phys.~Rev.~C~\VYP{#1}{#2}{#3}}
\def\PRD#1#2#3{Phys.~Rev.~D~\VYP{#1}{#2}{#3}}
\def\PRL#1#2#3{Phys.~Rev.~Lett.~\VYP{#1}{#2}{#3}}
\def\FBS#1#2#3{Few-Body~Sys.~\VYP{#1}{#2}{#3}}
\def\AP#1#2#3{Ann.~of Phys.~\VYP{#1}{#2}{#3}}
\def\ZP#1#2#3{Z.\ Phys.\  \VYP{#1}{#2}{#3}}
\def\ZPA#1#2#3{Z.\ Phys.\ A\VYP{#1}{#2}{#3}}
\def\half{\mbox{\small{$\frac{1}{2}$}}}
\def\quarter{\mbox{\small{$\frac{1}{4}$}}}
\def\nn{\nonumber}
\newlength{\PicSize}
\newlength{\FormulaWidth}
\newlength{\DiagramWidth}
\newcommand{\vslash}[1]{#1 \hspace{-0.42 em} /}
\def\olaf{\marginpar{Mod-Olaf}}
\def\her{\marginpar{$\Longleftarrow$}}
\def\bel{\marginpar{$\Downarrow$}}
\def\abo{\marginpar{$\Uparrow$}}

%\tighten

%%%%%%%%%%%%%%%%%%  TITLE, AFFILIATIONS, ABSTRACT  %%%%%%%%%%%%%%%%%%%%%

\title{Ratio of the proton electromagnetic 
form factors from meson dressing}

\author{S. Kondratyuk}
\affiliation{Department of Physics and Astronomy,
University of South Carolina, Columbia, SC 29208, USA 
\\and\\
Department of Physics and Astronomy, University of Manitoba,
Winnipeg, MB, Canada R3T 2N2}
\author{K. Kubodera}
\affiliation{Department of Physics and Astronomy,
University of South Carolina, Columbia, SC 29208, USA}
\author{F. Myhrer}
\affiliation{Department of Physics and Astronomy,
University of South Carolina, Columbia, SC 29208, USA}

\date{\today}

%\maketitle

\begin{abstract}
The Dressed K-matrix Model, developed previously for low- and intermediate-energy
Compton scattering, is generalized to calculate the photon-nucleon vertex with a virtual photon 
and an off-shell nucleon. Using this model, the ratio of the proton electromagnetic 
form factors is computed. The calculated ratio is in excellent agreement with the
ratio measured in polarized electron-proton scattering.
\end{abstract}

%Electromagnetic form factors
%Other models of strong interactions
%S-matrix theory; analytic structure of amplitudes
%Elastic electron scattering
%Elastic and Compton scattering

\maketitle

%%%%%%%%%%%%%%%%%  TEXT  %%%%%%%%%%%%%%%%%%%%%%%%%%%%

\section{Introduction} \seclab{intro}

Results of the recent experiments~\cite{Jon00} on polarized electron-proton scattering,
and their apparent disagreement~\cite{Arr03} with the electromagnetic form factors (EM FFs)
extracted from Rosenbluth cross section measurements, 
indicate that much remains to be learned about nucleon EM interactions.
While a definitive understanding of the nucleon EM FFs is still lacking,
several theoretical models~\cite{Fra96,Hol96,Lom01,Kas03,Dub04,Bij04} 
are able to describe their momentum-transfer dependence
consistently with the JLab experiments~\cite{Jon00}.
 
Since the ultimate goal is to reach a coherent understanding of all aspects of the 
nucleon EM interactions, important insights can be gained by 
describing as many different photon-nucleon reactions as possible in a unified 
theoretical approach. The Dressed K-matrix Model (DKM)~\cite{Kon00,Kon01,Kon02} 
has been developed for a description of real Compton scattering in a wide energy region 
both below and above the pion production threshold. The central element of this model
is a nonperturbative nucleon dressing based on the iterative use of dispersion relation, thus
adding analyticity constraints to the properties of relativistic covariance,
unitarity, crossing symmetry and gauge invariance of the usual K-matrix approaches.

The parameters of the model are not entirely free: the convergence requirement of the 
dressing procedure imposes constraints on the allowed range of these parameters while
introducing an interdependence among them. The thus constrained parameters were then 
completely fixed by a fit to Compton cross sections and pion-nucleon ($\pi N$) phase shifts
at intermediate energies~\cite{Kon01}.
With all the parameters fully fixed,  
low-energy observables--such as the nucleon polarizabilities~\cite{Kon01}, 
pion-nucleon scattering lengths and 
$\Sigma$-term~\cite{Kon02}--were then calculated and shown to agree with experiment. 
The main reason for this success is that important
analyticity (causality) constraints are implemented in DKM by the dressing of
the $\gamma N N$, $\pi N N$ and $\pi N \Delta$ vertices and of
the nucleon and $\Delta$ propagators with meson loops up to infinite order. 
The extent to which analyticity is
fulfilled can be quantified by comparing the low-energy amplitudes
with corresponding sum rules, both evaluated within the same model. Within DKM, 
such a comparison was
made for the Baldin-Lapidus and Gerasimov-Drell-Hearn sum rules vs.~the nucleon 
polarizabilities~\cite{KoS02}, and for the
Adler-Weisberger and Goldberger-Miyazawa-Oehme sum rules vs.~the near-threshold $\pi N$ 
amplitude~\cite{Kon04}. We generally found good agreement between the 
low-energy and sum-rule evaluations.  

In this report we extend the dressing procedure to the case of 
virtual space-like photons. The resulting dressed photon-nucleon ($\gamma N N$) vertex
has six invariant functions depending on two invariant variables: 
the four-momenta squared of the photon and of one of the nucleons, 
the other nucleon being on-shell. 
By putting both nucleons on the mass shell we calculate the momentum-transfer dependence 
of the ratio of the proton EM FFs. 
We find good agreement of our calculation
with the JLab polarization-transfer measurements~\cite{Jon00}. 

The crucial point of this result is that the 
momentum-transfer dependence of the ratio of the FFs is calculated
here with the the {\em same} parameters as used in Refs.~\cite{Kon02,Kon01,KoS02,Kon04} 
to describe the Compton and $\pi N$ amplitudes at low and intermediate energies. This shows that
the dynamics of DKM captures features which are important in a wide 
kinematical range relevant to the photon-nucleon interactions.

\section{Form factors in the dressed $\gamma N N$ vertex} \seclab{dress_eq}

The most general Lorentz-covariant structure of the $\gamma N N$ vertex with an incoming 
off-shell photon (four-momentum $q$) and an incoming off-shell nucleon 
(four-momentum $p$) can be written as~\cite{Bin60}
\beq
\Gamma^\mu(p,q)= \sum_{r=\pm} 
\left( \gamma^{\mu} F_1^{r}(p^2,q^2) +
i \frac{\sigma^{\mu \nu}q_{\nu}}{2m} F_2^{r}(p^2,q^2) +
\frac{q^{\mu}}{m} F_3^{r}(p^2,q^2) \right)\, \Lambda_r(\vslash{p}) \,,
\eqlab{vert_struct}
\eeq
where $\Lambda_{\pm}(\vslash{p})=(\pm \vslash{p} + m)/(2m)$ 
are the nucleon positive- and negative-energy
projection operators,\footnote{Throughout the paper 
we use the conventions and definitions of Ref.~\cite{Bjo64}.}
and the six invariant functions $F_i^{r}(p^2,q^2)$ ($i=\{1,2,3\}$, $r=\pm$) are called 
half-off-shell FFs. 
The outgoing nucleon is on the mass shell, i.~e.~its four-momentum
$p'=p+q$ obeys $p^{\prime\, 2}=m^2$. We consider protons interacting with real or
virtual space-like photons, i.~e.~$q^2 \le 0$. 

The usual Dirac and Pauli FFs, denoted as $F_D(q^2)$ and 
$F_P(q^2)$, respectively, 
are obtained by putting both the incoming and outgoing nucleons on the mass shell in \eqref{vert_struct}:
\beq
F_{D,P}(q^2) = \lim_{p^2 \rightarrow m^2} F_{1,2}^+(p^2,q^2).
\eqlab{diracpauli}
\eeq
When the two nucleons are on-shell,
$F_{1,2,3}^-(m^2,q^2)$ do not enter the vertex 
since $\Lambda_{-}(\vslash{p} = m)=0$; in addition, 
space-time reflection invariance requires~\cite{Bin60} that
\beq
F_3^+(m^2,q^2)=0.
\eqlab{pt_inv_f3pl}
\eeq
$F_{D,P}(q^2)$ are related to the Sachs (electric and magnetic) FFs $G_{E,M}(q^2)$ by
\beq
G_E(q^2)=F_D(q^2) + {{q^2}\over{4 m^2}} F_P(q^2), \;\;\;\;
G_M(q^2)=F_D(q^2) + F_P(q^2).
\eqlab{sachs}
\eeq 

The $\gamma N N$ vertex \eqref{vert_struct} is calculated by dressing 
a bare vertex with an infinite number of meson loops including pions
and vector isoscalar mesons. The $\pi N N$ and $\pi N \Delta$ vertices
and the nucleon and $\Delta$ propagators, which enter in the loop
integrals for the $\gamma N N$ vertex, are also dressed nonperturbatively
with pions, $\rho$ and $\sigma$ mesons.
The detailed description of the dressing technique for the $\gamma N N$ vertex (with real photons)
can be found in
Ref.~\cite{Kon00}, and for the $\pi N N$, $\pi N \Delta$ vertices 
and the propagators in Ref.~\cite{Kon02}. 
This dressing is part of DKM~\cite{Kon01} where it amounts to restoring the
principal-value parts of loop contributions to the Compton and $\pi N$ amplitudes.

In the present work we have generalized the dressing procedure of Ref.~\cite{Kon00} 
from real to virtual space-like photons in \eqref{vert_struct}. Thus
we have calculated the six invariant functions 
$F_i^{r}(p^2,q^2)$ for $-\infty < p^2 < \infty$ and $q^2 \le 0$,
from which the momentum-transfer dependence of the EM FFs has been obtained according to
Eqs.~(\ref{eq:diracpauli}) and (\ref{eq:sachs}).
For brevity, in this report we will focus only on the important new features 
related to the off-shellness of the photon.

We write the bare vertex as
\beq
\gamma^{\mu} +
i \frac{\sigma^{\mu \nu}q_{\nu}}{2m} \kappa_B +
\frac{q^{\mu}}{m} h_B(q^2).
\eqlab{vert_bare}
\eeq
The renormalization consists in choosing the bare constant $\kappa_B$ 
so that the dressed vertex with all particles on-shell reproduces the 
physical anomalous magnetic moment of the proton, 
i.~e.~$F^+_2(m^2,0)=\kappa_p = \mu_p -1 =1.79$, 
and adjusting 
$h_B(q^2)$ so that the calculated function $F_3^+(p^2,q^2)$ obeys \eqref{pt_inv_f3pl}.

\section{Gauge invariance of the model} \seclab{wti}

The Ward-Takahashi identity (WTI) is a consequence of gauge invariance of the theory~\cite{War50}.
It relates the $\gamma N N$ vertex \eqref{vert_struct} to the 
propagator $S(\vslash{p})$ of the off-shell proton: $q_\mu \Gamma^\mu(p,q) = -S^{-1}(\vslash{p})$. 
Thus the WTI dictates that
the half-off-shell FFs in \eqref{vert_struct} must obey the following relations:
\bea
F_1^+ (p^2,q^2) &=& \frac{m}{p^2-m^2} \left( 
\frac{\alpha(p^2)(p^2+m^2)}{m} - 2 \alpha(p^2) \xi(p^2) \right. \nn \\
&& \left. + {{q^2}\over{2m^3}} \left[(3m^2+p^2) F_3^+(p^2,q^2) +
(m^2-p^2) F_3^-(p^2,q^2) \right] \right),
\eqlab{wti_f1pl_off} \\
F_1^-(p^2,q^2) &=& \alpha(p^2) +{{q^2}\over{2m^2}} 
\left[ F_3^+(p^2,q^2) - F_3^-(p^2,q^2) \right],
\eqlab{wti_f1mi_off}
\eea
where the self-energy functions $\alpha(p^2)$ and $\xi(p^2)$ enter into 
the dressed nucleon propagator
$S(\vslash{p})={\ds \left[\alpha(p^2) (\vslash{p}-\xi(p^2)) \right]^{-1}}$
which was calculated together with the dressed $\Delta$ propagator and $\pi N N$
and $\pi N \Delta$ vertices~\cite{Kon02}.
The renormalization conditions $S^{-1}(m)=0$ and $\mbox{Res}\,S(m)=1$ can be explicitly written as
\beq
\alpha(m^2)\left(m-\xi(m^2)\right)=0,
\eqlab{ren_pole}
\eeq
\beq
2m \left(m-\xi(m^2) \right)
\frac{d \alpha(p^2)}{d p^2}\Bigg{\vert}_{p^2=m^2}-
\alpha(m^2)\left(2m\frac{d \xi(p^2)}{d p^2}\Bigg{\vert}_{p^2=m^2} -1 \right)=1.
\eqlab{ren_res}
\eeq

On expanding \eqref{wti_f1pl_off} in powers of $p^2$ around $m^2$ and 
using Eqs.~(\ref{eq:diracpauli}), (\ref{eq:pt_inv_f3pl}), (\ref{eq:ren_pole}) and 
(\ref{eq:ren_res}), we obtain the WTI for the Dirac FF of the proton:
\beq
F_D(q^2) = 1 + q^2 \left( 2 {{\pd F_3^+(p^2,q^2) \over \pd p^2}}
\Bigg{\vert}_{p^2=m^2} -
\frac{F_3^-(m^2,q^2)}{2m^2} \right),
\eqlab{wti_dir}
\eeq
yielding the familiar constraint $F_D(q^2=0) = 1$ for an on-shell vertex with a real photon.
Note that we do not impose Eqs.~(\ref{eq:wti_f1pl_off}) and (\ref{eq:wti_f1mi_off}) 
``by hand" at any stage of the dressing; nevertheless, the resulting
vertex does obey these constraints of the WTI due to gauge invariance of the model.

\section{Results of the calculation} \seclab{results}

The ratio of the calculated proton EM FFs is shown in 
\figref{ff_rat} as function of the photon momentum-squared, $Q^2 \equiv -q^2 \ge 0$,
together with the recent JLab measurements~\cite{Jon00}.
\begin{figure}[!htb]
\centerline{{\epsfxsize 15cm \epsffile[15 470 570 725]{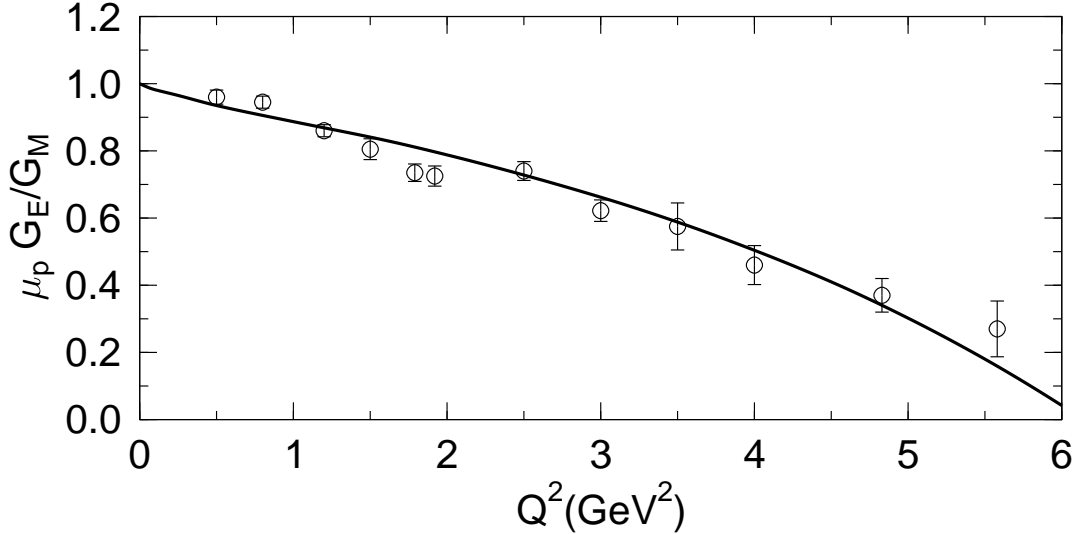}}}
\caption[f1]{Momentum-transfer dependence of the ratio of the proton EM FFs \eqref{sachs}
($Q^2 = -q^2$).
The present calculation is compared with the experimental results from polarized 
electron-proton scattering~\cite{Jon00}.
\figlab{ff_rat}}
\end{figure}
We found that, in order to obtain accurate agreement with experiment
over the wide range of momenta-transferred $Q^2$, 
the coupling of the vector isoscalar particle to the nucleon
has to be different from the $\omega N N$ coupling given in Ref.~\cite{Kon01}:
while the part of the vertex proportional to $\gamma_\mu$ is comparable to that of the $\omega N N$ vertex
(instead of $g_\omega=12$ as in \cite{Kon01}, now we use $g_{vect}=8$), the part proportional to 
$\sigma_{\mu \nu} q^\nu$ has to
be quite large (we need $\kappa_{vect}=60$ instead of 
$\kappa_\omega=-0.8$).\footnote{A large additional
vector particle contribution is also needed in 
other approaches in the literature, e.~g.~in a soliton model of Ref.~\cite{Hol96}.}
The other coupling constants are the {\em same} as in~\cite{Kon01}, i.~e.~as fixed in
the description of the $\pi N$ and Compton scattering processes. 
It is important that even without the contribution of the vector isoscalar particle, 
we find the slope of the ratio of the FFs to be negative at $Q^2=0$, although
in that case it is too steep. Including the vector particle in the dressing 
makes the slope less steep. 
A negative slope of the ratio implies that the
electric mean-square radius of the proton is larger that the magnetic 
mean-square radius normalized to the total magnetic moment, i.~e.~if
\beq
\left. \frac{d}{d Q^2}\frac{G_E(Q^2)}{G_M(Q^2)} \right|_{Q^2=0} < 0 ,
\eqlab{slope}
\eeq
then taking into account that $G_E(Q^2=0)=1$, $G_M(Q^2=0)=\mu_p\,$, and using the 
usual definitions of the mean-square radii
\beq
\left. \left< r_{E,M}^2 \right> = -6 \, \frac{d G_{E,M}(Q^2)}{d Q^2} \right|_{Q^2=0} ,
\eqlab{raddef}
\eeq
we find
\beq
\left< r_E^2 \right> > \frac{\left< r_M^2 \right>}{\mu_p} .
\eqlab{meanrads}
\eeq
Note that the traditional one-photon exchange
analyses of the Rosenbluth cross sections would yield $\mu_p \, G_E(Q^2)/G_M(Q^2)=const \approx 1$ 
(see, e.~g., Ref.~\cite{Arr03}), which would 
imply equal electric and (normalized) magnetic radii, in sharp contrast to \eqref{meanrads}.
We would also like to point out that the relation in \eqref{meanrads}
is consistent with the well-known fact that the electric polarizability of the proton, 
$\alpha_E \approx 12 \times 10^{-4} \mbox{fm}^3$, is larger 
than its magnetic polarizability,
$ \beta_M \approx 2 \times 10^{-4} \mbox{fm}^3$,
\beq
\alpha_E > \beta_M.
\eqlab{polaris}
\eeq
Thus one might speculate that the proton behaves as a ``larger" and hence more
``deformable" object in the presence of an electric field than in the presence of a magnetic field.
Since the mean-square radii are extracted from the momentum-transfer dependence of the
FFs while the polarizabilities from Compton scattering, 
the dressing procedure of DKM could provide a
theoretical framework for understanding 
this (possibly fortuitous) similarity between Eqs.~(\ref{eq:meanrads}) and (\ref{eq:polaris}).

Even though the ratio of the calculated FFs agrees with experiment very well
as shown in \figref{ff_rat}, the model fails in
describing $G_E(Q^2)$ and $G_M(Q^2)$ separately. This is demonstrated in \figref{ge_ff} for
the electric FF.
\begin{figure}[!htb]
\centerline{{\epsfxsize 15cm \epsffile[15 470 570 725]{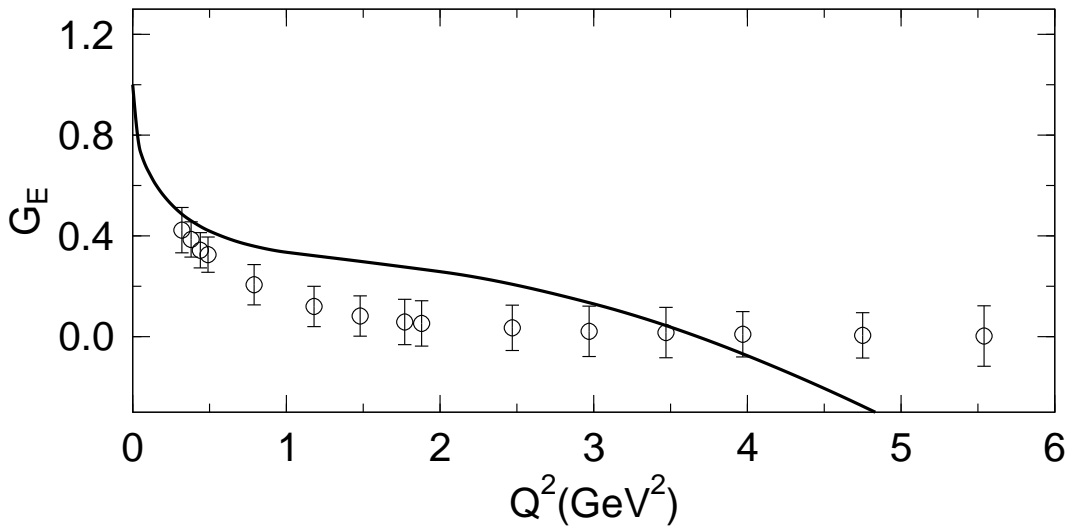}}}
\caption[f2]{The calculated electric FF of the proton is compared with 
the data analysis of Ref.~\cite{Arr04}.
\figlab{ge_ff}}
\end{figure}
A possible remedy could be to use an appropriate FF in the bare 
$\gamma N N$ vertex, which would introduce additional cut-off parameters. 
We have not done so in the present calculation since our aim is to keep the same model parameters
as fixed in describing Compton and $\pi N$ scattering. In this way we impose a stringent dynamical
constraint on the model.

The dependence of the calculated $\gamma N N$ 
vertex on the momentum-squared $p^2$ of the off-shell proton is illustrated
in \figref{gem_pdep}, where we show
the $p^2$-dependence of the ratio of generalized (half-off-shell) Sachs FFs 
which we define as
\beq
G_E^{gen}(p^2,q^2)=F_1^+(p^2,q^2) + {{q^2}\over{4 m^2}} F_2^+(p^2,q^2), \;\;\;\;
G_M^{gen}(p^2,q^2)=F_1^+(p^2,q^2) + F_2^+(p^2,q^2).
\eqlab{sachs_gen}
\eeq 
The usual Sachs FFs Eqs.~(\ref{eq:sachs}) are obtained from Eqs.~(\ref{eq:sachs_gen}) by
taking the limit $p^2 \rightarrow m^2$ in accordance with \eqref{diracpauli}. 
\begin{figure}[!htb]
\centerline{{\epsfxsize 15cm \epsffile[15 150 570 450]{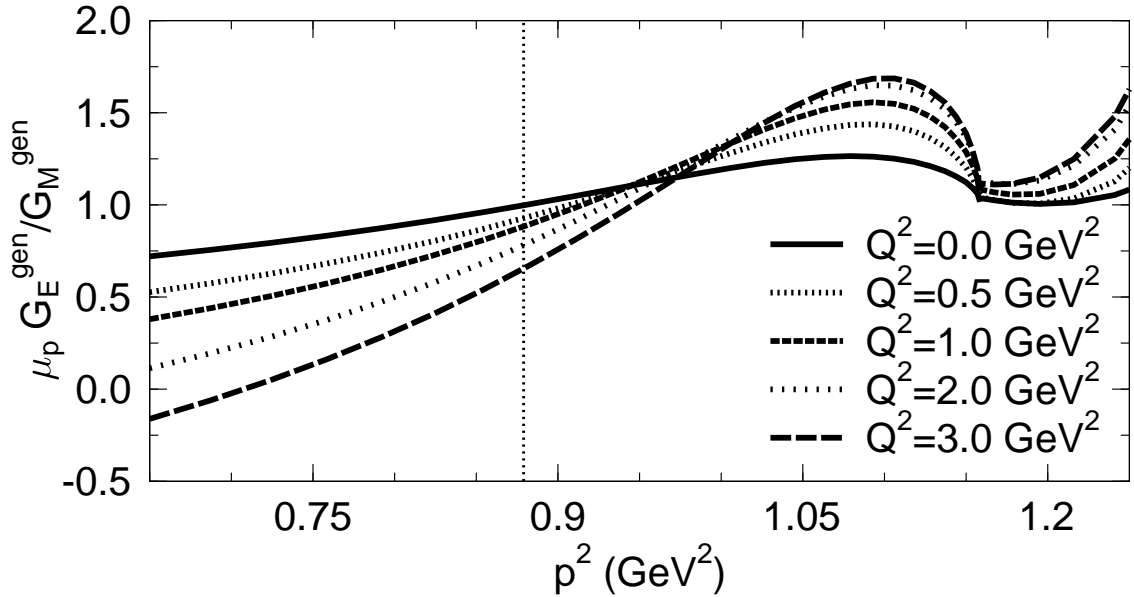}}}
\caption[f3]{Ratio of the generalized Sachs form factors Eqs.~(\ref{eq:sachs_gen})
as function of the momentum-squared of the off-shell proton, for various fixed
values of the photon momentum-squared $Q^2 = -q^2$. The vertical dotted
line corresponds to both nucleons being on-shell with $p^2=p^{\prime 2}=m^2$. 
The cusp at $p^2 \approx 1.16$ GeV$^2$ indicates the opening of the pion production threshold.
\figlab{gem_pdep}}
\end{figure}
Although the half-off-shell FFs are not measurable by themselves~\cite{Fea00}, 
they can be an important part of various models for physical processes.
For example, a two-photon exchange contribution to 
electron-proton scattering contains $\gamma N N$ vertices with an off-shell
nucleon leg. So far the dependence of these vertices on the
nucleon off-shell momentum has not been taken into account 
in the existing approaches~\cite{Blu03}.
Our calculation shown in \figref{gem_pdep} suggests that this dependence may be significant.
It should be borne in mind, however, that in view of the representation-dependence of 
the Green's functions~\cite{Chi61}, 
the calculation of such off-shell vertices should be consistent with
the model for the physical processes to which they are applied.

\section{Conclusions} \seclab{concl}

In this report we extended the nonperturbative
dressing procedure of the Dressed K-matrix Model (DKM)
to calculate a $\gamma N N$ vertex with 
an off-shell nucleon and an off-shell photon. The principal motivation for 
this work is to study the nucleon electromagnetic interactions in a dynamical 
approach which describes in a
unified manner two distinct types of reactions: those where essential contributions come from
an off-shell nucleon (as in Compton scattering) and those where one probes the dependence 
on the momentum transferred by an off-shell photon (as in form factors 
extracted from electron-proton scattering experiments). 
Having previously applied DKM to Compton and $\pi N$ scattering,
in this report we focused on the momentum-transfer dependence. We developed 
the dressing procedure to calculate the ratio of the proton electromagnetic form factors and found
good agreement of our results with the recent JLab measurements. 

The main limitation of the present version of DKM is the difficulty in describing
the momentum-transfer dependence of the individual form factors as accurately as their ratio.
To resolve this problem, additional dynamical contents--with attendant new
parameters--might be required in the model. Rather than pursuing this direction in detail,
we studied here to what extent one can describe
the dependence of the nucleon EM interactions on both the nucleon and photon variables 
using the {\em same} dynamical approach with a few predetermined parameters. 
Our approach is based on essential symmetry constraints, including
relativistic and gauge invariance, unitarity, crossing and causality, which are
required in order to correlate in one model the very different
kinematical regions explored in Compton and $\pi N$ scattering and in the proton form factors.

\begin{acknowledgments}

We thank Shmuel Nussinov, Ralf Gothe, Pawel Mazur and Peter Blunden for useful discussions.
This work was supported in part by the US National Science Foundation's
Grant No.~PHY-0140214. One of the authors (FM) appreciates the support and
hospitality of the Nuclear Theory Group at Bonn University.

\end{acknowledgments}

%%%%%%%%%%%%%%%%%%%%%%%%%%%%%%%% BIBLIOGRAPHY %%%%%%%%%%%%%%%%%%%%%%%%%%%%%%

\end{document}